\documentclass[aps,reprint,pra,showpacs]{revtex4-1}
\usepackage{graphicx}% Include figure files
\usepackage{dcolumn}% Align table columns on decimal point
\usepackage{bm}% bold math
\usepackage{amsmath}
\usepackage{latexsym}
\usepackage{url}
\numberwithin{equation}{section}

\begin{document}
\def\bfr{{\bf r}}
\def\nbar{{\bar n}}

\title{Range optimized theory of electron liquids with application to the homogeneous gas}
\author{James P. Donley}
\email{jdonley@valence4.com}
\affiliation{Department of Materials and Metallurgical Engineering, New Mexico Institute of Mining and Technology, Socorro, NM, 87801}
\altaffiliation[Also at ]{Valence4 Technologies, Arlington, VA 22202}

\begin{abstract}A simple optimization scheme is used to compute the density-density response function of an electron liquid. Higher order terms in the perturbation expansion beyond the random phase approximation are summed approximately by enforcing the constraint that the spin density pair correlation functions be positive. The theory is applied to the 3-D homogeneous electron gas at zero temperature. Quantitative comparison is made with previous theory and data from quantum Monte Carlo simulation. When thermodynamic consistency is enforced on the compressibility, agreement with the available simulation data is very good for the entire paramagnetic region, from weakly to strongly correlated densities. In this case, the accuracy of the theory is comparable to or better than the best of previous theory, including the full GW approximation. In addition, it is found that the spin susceptibility diverges at a lower density ($r_s \approx 107$) than the current estimate for the liquid-solid transition. Application of the theory to inhomogeneous electron liquids is discussed.
\end{abstract}

\date{\today}

\maketitle

\section{\label{sec:intro}Introduction}
In many practical calculations of electronic structure, such as for semiconductors and molecules, the density-density response function $\chi$ plays a central role. The
popular Kohn-Sham electronic density functional theory (DFT)\cite{kohn99,perdew02,capelle06}, and Hedin's ``GW'' approximation (GWA) of many-body perturbation theory\cite{hedin65,aulbur00} are two such methods that use $\chi$.

In DFT, $\chi$ is an important ingredient in the
exchange-correlation functional, $E_{xc}$, which is the main object for
approximations in the theory.\cite{perdew02} The workhorse local density
approximation (LDA) and its generalizations avoid explicitly
estimating $\chi$ for the inhomogeneous liquid. Instead, they approximate
$E_{xc}$ by relying on knowing the correlations of a simpler system, the
homogeneous electron gas, i.e, jellium.\cite{giuliani05} The correlations of
this latter system have been computed accurately using theory and quantum
Monte Carlo (QMC) simulation.\cite{gorobchenko89} While being very accurate for
the electron structure of molecules and electron density of many systems, 
 DFT-LDA and its variants have the limitation of not being able to
describe accurately band gaps, nor London dispersion (i.e., van der Waals) 
interactions very well.\cite{perdew02} The former is crucial to understanding
the properties of semiconductors, while the latter is important to understanding
weakly bonded systems, such as water. A recent branch of DFT uses the
adiabatic-connection fluctuation-dissipation form for the correlation energy,
which is the troublesome part of $E_{xc}$. In this branch there is much current
interest in the use of 
the random phase approximation (RPA) for $\chi$.\cite{angyan11}
Unlike the LDA, this DFT-RPA does predict reasonable band gaps for many 
solids,\cite{niquet04} and, with some extra effort, reasonable London dispersion interactions between atoms.\cite{toulouse09,klimes12}

In GWA, $\chi$ enters through the screened potential $W$ to compute ultimately
the one-particle Green's function, $G$, from which the
band structure is extracted.\cite{hedin65,giuliani05} In perturbative calculations around a DFT reference state, known as ``one-shot GW,'' $\chi$ is almost always taken to have a simple RPA form. In the full version of GWA, $\chi$ is
computed in a self-consistent loop along with $G$.
Being able to predict accurate band structure for many solids, GWA
has rapidly become one of the main methods for computing the electronic structure of
crystalline materials.\cite{bruneval06,yu10} A drawback to this approach though is
that the full, self-consistent solution often does not improve the predictions of the
theory, and can make it worse.\cite{holm98} Application of the GWA to
jellium has shown that the main weakness is the expression for $\chi$ obtained
from $G$.\cite{holm98} 

For both approaches then, a better way to compute $\chi$ would be very desired.
In this paper, a simple scheme called range optimization is described that
goes beyond the RPA for $\chi$ to accomplish this task.
Range optimization was originally used to improve the
RPA theory of classical molecules. This ``range-optimized'' RPA (RO-RPA) theory
was able to describe very well the
equilibrium structure and thermodynamics of strongly charged polyelectrolyte
solutions.\cite{donley04,donley05,donley06} It has also been applied
successfully to the study of the hydrated electron.\cite{donley14}
Given that a number of weaknesses of the GWA became apparent only after the
full version had been implemented for jellium, it is important for the theory
here to be analyzed first for that system. This is the intent of the
present paper.  As will be
shown below, the scheme applied to jellium greatly improves the predictions of
the theory over the basic RPA. A new algorithm to implement range optimization,
valid for inhomogeneous liquids, is also described here.

\section{\label{sec:theory}Theory}
Let the analysis be confined to a homogeneous electron gas, i.e., jellium, in
3-D at zero temperature in the paramagnetic phase. As a reminder, jellium
has a balancing background of positive charge that is uniform and rigid.\cite{fetter03,giuliani05}

The theory here is centered around the
time-ordered, spin density-density response function (or spin polarization
propagator), $\chi_{ij}(x,x')$,
where $i$ and $j$ label the spin ($\uparrow$ and $\downarrow$), and
$x\equiv \{\bfr,t\}$, with $\bfr$ being the position and $t$ the
time. It is defined as:
\begin{equation}
\chi_{ij}(x,x') ={1\over i\hbar}{\langle\Psi_0\vert T[\delta{\hat n}_i(x)\delta{\hat n}_j(x^\prime)]\vert\Psi_0\rangle\over \langle\Psi_0\vert\Psi_0\rangle}.
\label{chidef}
\end{equation}
Here, $\vert\Psi_0\rangle$ is the system ground state, and $T$ denotes the
time ordered product. Also,
\begin{equation}
\delta{\hat n}_i(x) = {\hat n}_i(x) - \langle {\hat n}_i(x)\rangle,
\label{deltan}
\end{equation}
where ${\hat n}_i(x)$ is the number density operator in the Heisenberg 
representation for spin $i$ electrons at point $x$,\cite{fetter03} and
$\langle {\hat n}_i(x)\rangle$ denotes its ground state average.
For the paramagnetic phase, $\langle {\hat n}_\uparrow(x)\rangle = \langle {\hat n}_\downarrow(x)\rangle = n/2$, where $n$ is the average electron density.

Jellium is translationally invariant and $\chi$ is symmetric with respect to
$x\leftrightarrow x'$, so $\chi_{ij}(x,x')\rightarrow \chi_{ij}(r,\tau)$, where $r = \vert \bfr -\bfr'\vert$ and $\tau = t - t'$. 
For this case, it is helpful to work with the dual Fourier transform,
$\chi_{ij}(k,\omega)$, $k = \vert {\bf k}\vert$ being the wavevector and
$\omega$ the frequency.

A number of useful quantities can be obtained from $\chi_{ij}(k,\omega)$.\cite{giuliani05} First is the system compressibility $K$:
\begin{equation}
{K_0\over K} = - N(0)\lim_{k\rightarrow 0}\left[ {1\over \chi(k,0)} + v(k)\right],
\label{compresseq}
\end{equation}
where $\chi(k,\omega) =\sum_{ij}\chi_{ij}(k,\omega)$ is the total 
density-density response function, and $N(0) = m k_F/(\pi^2\hbar^2)$, with $m$
being the electron mass, and $k_F = (3\pi^2 n)^{1/3}$ being the Fermi
wavevector.\cite{giuliani05} Also, $K_0$ is the
compressibility of the non-interacting gas, and $v(k)$ is the Fourier transform
of the electron-electron Coulomb potential, $v(r) = e^2/r$, with $e$ being the
electron charge. Second is the spin susceptibility $\chi_S$:
\begin{equation}
{\chi_P\over \chi_S} = - N(0)\lim_{k\rightarrow 0}\left[
{1 \over 2(\chi_{\uparrow\uparrow}(k,0) - \chi_{\uparrow\downarrow}(k,0))}\right],
\label{suscepteq}
\end{equation}
where $\chi_P$ is the spin susceptibility of the non-interacting gas.
Third are the partial static structure factors:
\begin{equation}
S_{ij}(k) = {2i\hbar\over \pi n}\int_{0}^{\infty} d\omega
\chi_{ij}(k,\omega),
\label{skeq}
\end{equation}
which are real, and have exploited that $\chi_{ij}(k,\omega)$ is symmetric
about $\omega=0$.
Fourth are the spin-spin pair correlation (or radial distribution) functions:
\begin{equation}
g_{ij}(r) = 1 + {2\over n}\int{d{\bf k}\over (2\pi)^3} 
\bigl [ S_{ij}(k) - \delta_{ij}\bigr ] exp(i{\bf k}\cdot{\bf r}),
\label{greq}
\end{equation}
where $\delta_{ij}$ is the Kronecker delta. The pair correlation function
$g_{ij}(r)$ is proportional to the equilibrium probability density of there
being an electron of spin $j$ a distance $r$ from one with spin $i$ at the same
time. As a density, $g_{ij}(r)$ is strictly positive for all $r$.
Last, the correlation energy $E_{corr}$ can be obtained
from $S_{ij}(k)$.\cite{fetter03} Two expressions for $E_{corr}$ were used
and they are given in Sec. \ref{sec:numerical} below. 

Define the matrix inverse of $\chi_{ij}(k,\omega)$ by
$\sum_s \chi_{is}(k,\omega)\chi_{sj}^{-1}(k,\omega) =\delta_{ij}$.
This inverse can be represented exactly as:
\begin{equation}
\chi_{ij}^{-1}(k,\omega) = {\tilde\chi}_{ij}^{-1}(k,\omega) - v_{ij}(k),
\label{chiinverse}
\end{equation}
where $v_{ij}(k) = v(k)$, and
${\tilde\chi}_{ij}(k,\omega)$ is the proper spin density-density response
function (or proper spin polarization propagator).\cite{fetter03,giuliani05}
It is helpful to rewrite Eq.(\ref{chiinverse}) as:
\begin{equation}
\chi_{ij}^{-1}(k,\omega) = \chi_{0ij}^{-1}(k,\omega) -
 v_{ij}(k) - u_{ij}(k,\omega),
\label{chiinverse2}
\end{equation}
where $u_{ij}(k,\omega) = \chi_{0ij}^{-1}(k,\omega) -
{\tilde\chi}_{ij}^{-1}(k,\omega)$ and 
$\chi_{0ij}(k,\omega) = \delta_{ij}\chi_0(k,\omega)/2$, with $\chi_0(k,\omega)$
being the total density-density response function of the non-interacting gas,
i.e., the Lindhard function. This Lindhard function can be computed analytically.\cite{fetter03}

Setting $u_{ij}(k,\omega)$ to zero reduces Eq.(\ref{chiinverse2}) to
the familiar RPA expression for $\chi_{ij}$.\cite{giuliani05} As is well known,
the RPA tends to work well if the interactions are weak.
However, for strongly interacting systems it works less well, causing, for
example, the pair correlation functions $g_{ij}(r)$ to be (very) negative at
small $r$.

More recent research on jellium has steadily improved upon the RPA.
Almost all of these theories have worked with the one-component
expression for the total density-density response function, $\chi(k,\omega)$,
by developing accurate approximations to the
static and dynamic local field factors, $G_+(k)$ and $G_+(k,\omega)$,
respectively. These are defined by: 
\begin{equation}
\chi^{-1}(k,\omega) = \chi_0^{-1}(k,\omega) - v(k)[1 - G_+],
\label{gfactoreq}
\end{equation}
where $G_+$ denotes $G_+(k)$ or $G_+(k,\omega)$. In this manner,
the one-component optimized potential, $u = -vG_+$.

 The best of these theories now agree with QMC simulation data for the paramagnetic state for the correlation energy $E_{corr}$
within a few percent for the density range of most metals, $2 < r_s < 6$.\cite{vosko80,gorobchenko89}
Here, $r_s = r_0/a_0 = 1/(\alpha k_F a_0)$, where $r_0 = ({3/(4\pi n)})^{1/3}$
is the average distance between electrons, $a_0$ is the Bohr radius, and
$\alpha = (4/(9\pi))^{1/3}$.
A limitation of these theories though is that they usually apply only
to the paramagnetic, i.e., zero polarization state.
This local field factor approach can be extended to examine partially polarized
states, and thus give information about the jellium phase diagram.\cite{tanaka89}
However, the cost is an increase in the complexity of the theory.
As such, a different path will be taken here.

To go beyond the RPA for the multi-component model, Eq.(\ref{chiinverse2}), the
range optimization scheme will be used. This scheme has been described in
detail elsewhere,\cite{donley04,donley05} but a summary is given here.

The aim is to approximate
the higher order terms embodied in $u_{ij}(k,\omega)$ in some manner.
First, let $u_{ij}(k,\omega)$ be independent of frequency $\omega$. This approximation is not necessary, but is a sensible one for computing the static
equilibrium properties of the gas, such as $g_{ij}(r)$ and $E_{corr}$. Next,
let $u_{ij}(k)$ be real. This follows the common assumption that the static
local field factor $G_+(k)$ is dominated by its real part.\cite{giuliani05}
In this manner, the inverse Fourier transform of $u_{ij}(k)$, $u_{ij}(r)$, can
be viewed as a short-ranged attractive potential that
counteracts the strong electron-electron Coulomb repulsion in $v(r)$ at
small $r$. What then is $u_{ij}(r)$?

Now, the RPA works well at high density, $r_s < 1$, where the
kinetic energy and exchange interactions dominate the Coulomb repulsion. At
low density, $r_s \gg 1$, where the RPA breaks down, the electron-electron
Coulomb repulsion causes $g_{ij}(r)$ to be essentially zero out to some
range $\sigma_{ij}$ (see Figure 1). Notice though that if $g_{ij}(r)$ were zero,
it makes little difference to the electrons that the repulsion at that distance
were infinite as opposed to just very large. As such, replace the Coulomb
potential with a hard-core one for distances $r<\sigma_{ij}$.  While the RPA
fails for hard-core potentials naturally, methods developed in the
classical theory of liquids have found ways to overcome this problem.\cite{hansen86} One is the
mean spherical approximation (MSA) closure.\cite{lebowitz66}

Applied to jellium, the MSA states that if $\sigma_{ij}$ is the range of
the hard-core potential between electrons of spin $i$ and $j$, then $g_{ij}(r)$ is zero inside this range. But since
$g_{ij}(r)$ is zero inside, the equations relating $u_{ij}(r)$ to
$g_{ij}(r)$ can be used to determine $u_{ij}(r)$ inside. That is,
$u_{ij}(r)$ takes whatever form is needed to ensure that $g_{ij}(r)$ is
zero inside the core. The closure is summarized as:
\begin{eqnarray}
g_{ij}(r) && = 0, \ r < \sigma_{ij}, \nonumber \\
u_{ij}(r) &&=0, \ r > \sigma_{ij}.
\label{msaclosure}
\end{eqnarray}
This closure, along with Eqs. (\ref{skeq}), (\ref{greq}) and 
(\ref{chiinverse2}), form a closed set of equations for $g_{ij}(r)$ and
$u_{ij}(r)$, assuming the $\sigma_{ij}$ are known. The last step is
to optimize the range, $\sigma_{ij}$,
by letting it have the smallest value such that $g_{ij}(r)$ is
positive for all $r$. Since the theory will now work properly for low and
high densities, it presumably will work well for intermediate densities as well.
This set of self-consistent equations will be referred to as RO-RPA theory.

There are at least two ways to compute the compressibility: by the
structure route using $\chi(k,0)$ in Eq.(\ref{compresseq}), and by
the energy route using an expression for the total energy\cite{giuliani05}.
Since the RO-RPA theory is approximate, the structure and energy routes will
not give the same value for $K$. It is well known though that enforcing 
consistency between these two routes, i.e., enforcing the ``sum rule'', can
improve a theory, often greatly.\cite{dickman96} This
thermodynamic consistency can be attained under range optimization by
noticing that $g_{ij}(r)$ need not be set to zero inside the core. Instead, the minimum of $g_{ij}(r)$ could have a non-zero value,
$g_0$. The MSA closure can then be generalized to:
\begin{eqnarray}
g_{ij}(r) && = g_0, \ r < \sigma_{ij}, \nonumber \\
u_{ij}(r) &&=0, \ r > \sigma_{ij},
\label{msaclosure_tc}
\end{eqnarray}
where the value of $g_0$ is determined by enforcing the sum rule on the
compressibility. This extra condition, along with the RO-RPA equations above,
will be referred to as ``thermodynamically consistent'' RO-RPA (TCRO-RPA) theory.  

\section{\label{sec:numerical}Numerical solution}
All theories were solved numerically as follows.
Functions of $r$ or $k$ were solved on a grid of $N_r$
points with spacing $\Delta r$ or $\Delta k = \pi/(N_r\Delta r)$,
respectively. Unless stated otherwise, $N_r$ and $\Delta r$ were set to
$2^{11}$ and $0.05/k_F$, respectively. To compute the static structure factor
 $S_{ij}(k)$, the integration of $\chi_{ij}(k,\omega)$ over $\omega$ given
by Eq.(\ref{skeq}) was performed
along the positive real axis. It is well known that along the real axis,
a contribution to $S_{ij}(k)$ from the plasmon mode must be accounted
for\cite{pines99,giuliani05} and that was done.
As a check though, the integration was also performed along the positive 
imaginary axis.\cite{pines99,hedin65} In either case, the integral over
frequency was evaluated using Romberg integration\cite{numrecipes}
with the relative error tolerance being $10^{-6}$ and
$10^{-9}$ for the real and imaginary axis cases, respectively. 

Once $S_{ij}(k)$ was computed, the integral over $k$ in Eq.(\ref{greq}) was
evaluated by inverse Fourier transform to obtain $g_{ij}(r)$. As a check on the
accuracy, the RPA values for $g_{ij}(0)$ were computed. It was found that
unlike the correlation energy, $E_{corr}$, $g_{ij}(0)$ was sensitive to the
grid spacing, but setting
$N_r=2^{16}$ and $\Delta r = 0.05/(32k_F)$ gave convergent RPA values of
$g_{ij}(0)$ within $0.1\%$.\footnote{The RPA values for the spin-averaged
$g(0)$ agree best with those of Toigo and Woodruff\cite{toigo71} and Gorobchenko et al.\cite{gorobchenko89}} The grid spacing did not affect greatly any other
quantity, though for increased accuracy in determining the density at which the
spin susceptibility $\chi_S$ diverged, $N_r$ and $\Delta r$ were set to
$2^{13}$ and $0.0125/k_F$, respectively, for $r_s \geq 70$.

A new algorithm was used to implement range optimization.
This new algorithm has two advantages over the one used in past
work\cite{donley04,donley05,donley06,donley14}: it is straightforward to apply
to inhomogeneous liquids, and is more efficient even for jellium.
It is as follows: 1) An initial guess is made for the optimized potentials 
$u_{ij}(r)$, which could be zero. 2) With the $u_{ij}(r)$, the theory is solved
as described above for the pair correlation functions $g_{ij}(r)$. 3) The
difference $\Delta g_{ij}(r) = g_{ij}(r)-g0$ is computed for all $r$, where
$g0=0$ for standard range optimization. 4) As a variation on Picard
iteration\cite{hansen86}, the change in the value of the optimized potential
is set to $\alpha \Delta g_{ij}(r)$, with the mixing parameter
$\alpha \leq 0.25$. 5) Since the optimized potential is an attractive
potential, its new value for each $r$ is checked to determine
if it is greater than zero; if so, it is set to zero. 6) The difference
between the new and old values is checked to determine if the potential
has converged; if not, the steps starting at 2) are repeated until convergence
is obtained. Here, the relative error tolerance on each point of $u_{ij}(r)$ was
$10^{-4}$, although in some cases it was reduced to $10^{-5}$ as a check.

Note that in this algorithm the optimized ranges, $\sigma_{ij}$, are not
considered explicitly. Instead, the algorithm relies on knowing only  
the value of the pair correlation function, $g_{ij}$, to obtain a refined guess for the optimized potential, $u_{ij}$, at the same point. In that way, the
above algorithm can be used for inhomogeneous liquids by the mere replacement of
$g_{ij}(r)$ and $u_{ij}(r)$ with $g_{ij}(\bfr_1,\bfr_2)$ and $u_{ij}(\bfr_1,\bfr_2)$, respectively, and then using the inhomogeneous analog of
Eq.(\ref{chiinverse2}).\cite{martin04}

For all theories except TCRO-RPA the correlation energy was computed in the
usual manner as a charging integral over the coupling constant $e^2$, i.e.,
via the adiabatic-connection fluctuation-dissipation theorem.\cite{fetter03}
Define $\epsilon_c = 2a_0E_{corr}/(Ne^2)$ as a scaled correlation energy per 
particle (in units of Rydbergs), with $N$ being total number of electrons.
Then this energy equation is:
\begin{equation}
\epsilon_c(r_s) = {4\pi\over \alpha r_s}\int_0^1 d\lambda\ w(\lambda),
\label{eq:correnergy1}
\end{equation}
where
\begin{equation}
w(\lambda) ={1\over 2\pi^2k_F}\int_0^\infty dk \left[S(k,\lambda) - S(k,0)\right ].
\label{eq:woflambda}
\end{equation}
Here, $S(k,\lambda) = {1/2}\sum_{ij}S_{ij}(k,\lambda)$ is the total structure
factor for a gas with electron-electron potential $\lambda v(r)$.
The integral over $k$, Eq.(\ref{eq:woflambda}), was computed via quadrature.
The charging integral, Eq.(\ref{eq:correnergy1}), was computed using Romberg
integration with an error tolerance of $10^{-4}$.

Enforcing thermodynamic consistency for the TCRO-RPA theory was done as follows.
First, for a given value of $r_s$, a value for $g0$ was guessed. Then the
optimized potentials, $u_{ij}(r)$, were computed in the same manner as for
the RO-RPA. Next, the compressibility was computed using the structure route
formula, Eq.(\ref{compresseq}). For the compressibility via the energy route, its value
was computed initially using the Perdew-Wang fit for the correlation energy
$E_{corr}$,\cite{perdew92} and an expression relating the total energy to the
compressibility.\cite{giuliani05}
The change in the value of $g0$ was set proportional to the difference between
values of $K_0/K$ from these two routes. With this new $g0$, steps 2)-6) above
were repeated, and this iteration was continued until the value of $g0$
converged. This procedure was repeated to obtain $\chi_{ij}(k,\omega)$ on a grid
for $0\leq r_s\leq 11$ (see below) with spacing $\Delta r_s=0.1$.

These density-density response functions
were then used to compute the correlation energy over this range.
The representation of $E_{corr}$ expressed as an integral over $r_s$ was
used:\cite{stls68,utsumi80}
\begin{equation}
\epsilon_c(r_s) = \epsilon_c^{rpa}(r_s) - {4\over\pi\alpha r_s^2}\int_0^{r_s}dx \Delta\gamma(x),
\label{ecorreq}
\end{equation}
where
\begin{equation}
\Delta\gamma(r_s) =  -{1\over 2k_F}\int_0^\infty dk \left [
S(k,r_s) - S_{rpa}(k,r_s)\right ].
\label{gammaeq}
\end{equation}
Here, $\epsilon_c^{rpa}(r_s)$ and $S_{rpa}(k,r_s)$ are the RPA values for
the correlation energy and total structure factor, respectively, at mean 
separation $r_s$.
An accurate interpolation formula for $\epsilon_c^{rpa}$ due to
Perdew-Wang\cite{perdew92} was used here.

The integral, Eq.(\ref{gammaeq}), for $\Delta\gamma(r_s)$ was 
computed for each $r_s$ in the same manner as for Eq.(\ref{eq:woflambda}).
The energy route expression for the compressibility consists of
derivatives of $\epsilon_c$ with
respect to $r_s$. To minimize errors then, this set of $\Delta\gamma$ values was
then fit to an $n$th degree polynomial in $r_s$. Degree $n=11$ was found to give
a sufficiently accurate fit ($n=9$ worked almost as well).
With this functional form, Eq.(\ref{ecorreq}) was evaluated analytically.
The self-consistent theory for $g_0$ was then solved again, with the new
and old values for $\epsilon_c$ being used to compute the energy route $K$
with a mixture of 1:1 old to new. After new density-density
response functions, $\chi_{ij}(k,\omega)$ were computed, the procedure to
compute new values for $\epsilon_c$ was repeated. It was found that the
fitted values for $K$ had converged to within $10^{-3}$ ($10^{-5}$ for
$\epsilon_c$) for $r_s\leq 10$ after seven iterations.

At $r_s\approx 0$, $g_0\approx 0$ naturally, then $g_0$ rose to a maximum of
0.177 at $r_s\approx 1.7$, and then gradually dropped to zero again at
$r_s\approx 10.8$. When the positivity constraint on $g_{ij}(r)$ was relaxed,
$g_0$ became slightly negative as $r_s$ increased beyond 10.8. Since this
positivity constraint on $g_{ij}(r)$ is more
important than enforcing a sum rule, $r_s\approx 10.8$ is then the limit of
the usefulness of enforcing thermodynamic consistency on the compressibility.
However, it will be shown below that since the RO-RPA is most
accurate at low density, this limit is not regarded as important.

For comparison, some results of the theories of Singwi-Tosi-Land-Sj{\"o}lander (STLS),\cite{stls68} 
and Utsumi and Ichimaru (UI) \cite{utsumi80} will also be shown. The UI theory is
considered accurate for the short range behavior of the pair correlation
function at metallic densities.\cite{gorobchenko89} The STLS theory is
considered accurate for the correlation energy
and almost as accurate as UI for the pair correlation function, but is
also straightforward to implement. The theory has also been generalized to apply to
inhomogeneous liquids in atoms and ions.\cite{gould12} The STLS theory was solved for
the total structure factor $S(k)$ in a similar manner to that of $S_{ij}(k)$
for the RO-RPA. Once $S(k)$ was determined self-consistently, the spin-averaged
$g(r)={1/4}\sum_{ij}g_{ij}(r)$ was obtained by inverse Fourier
transform using the analog of Eq.(\ref{greq}). The UI theory was solved in the
same manner as the RPA, but with $v(k) \rightarrow v(k)(1-G_+(k))$. Values for
the static local field factor $G_+(k)$ were
interpolated from data presented in Table I of \cite{utsumi80}.

\section{\label{sec:results}Results}
Unless noted otherwise, all RO-RPA results given here will be for the
multi-component version described in Sec.\ref{sec:theory} above.
Results of the one- and multi-component versions of the RPA are the same for
the quantities examined here.

\begin{figure}
\includegraphics[scale=0.72]{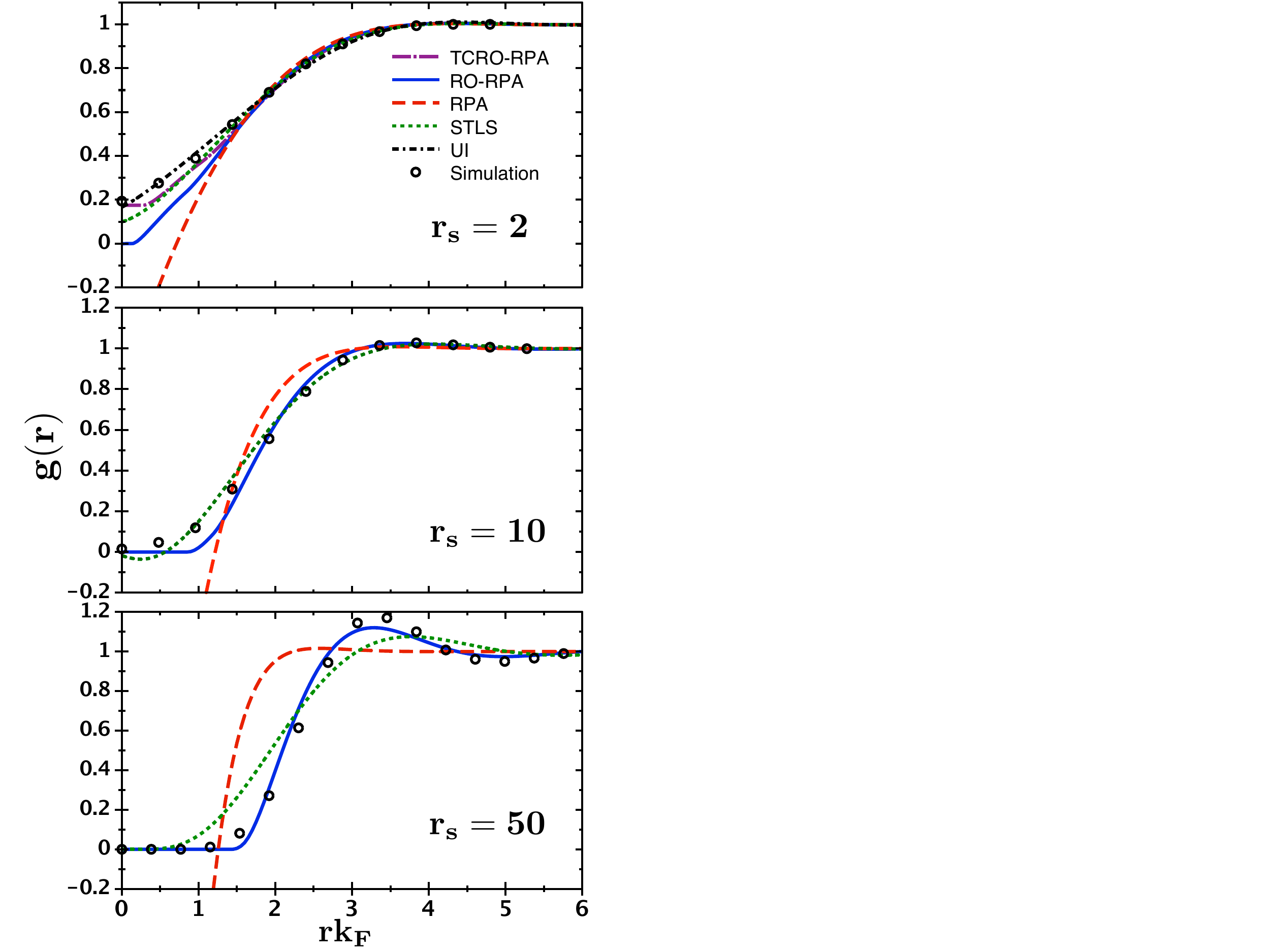}
\caption{\label{fig:gr}
Spin-averaged pair correlation function $g(r)$ as a function of distance $r$
for various densities.
The meanings of the curves and symbols are shown in the figure legend.
For $r_s=$ 2 and 10, the simulation data are
from Ceperley and Alder\cite{ceperley80} as obtained from Gori-Giorgi et 
al.\cite{gori00} For $r_s=50$, the simulation data are from
Zong et al.\cite{zong02} The contact value $g(0)$ from the RPA theory is
-0.66, -3.95 and -15.1 for $r_s=$ 2, 10 and 50, respectively.
}
\end{figure}

Figure \ref{fig:gr} shows results for the spin-averaged pair
correlation function, $g(r)$, for various
densities. As a comparison, QMC simulation data of
Ceperley and co-workers is also shown.\cite{ceperley80,zong02} As can be 
seen, the predictions of the RO-RPA are much improved over the RPA, with the
RO-RPA outperforming, as expected, the STLS theory at very low density,
$r_s=50$. For $r_s=2$, the contact value, $g(0)=0.176$ and 0.175 for the
UI and TCRO-RPA theories, respectively, which are 10\% smaller than the
simulation value.
For $r_s=10$, the thermodynamically consistent value of $g_0$ was close to
zero. Consequently, the TCRO-RPA prediction for $g(r)$ (not shown) is almost
the same as the RO-RPA for this density (and lower densities). Holm and 
von Barth have shown that the one-shot and fully self-consistent GWA produce
only modest improvement in the local structure of $g(r)$ over the RPA, for the
metallic density $r_s=4$.\cite{holm04}
\begin{figure}
\includegraphics[scale=0.56]{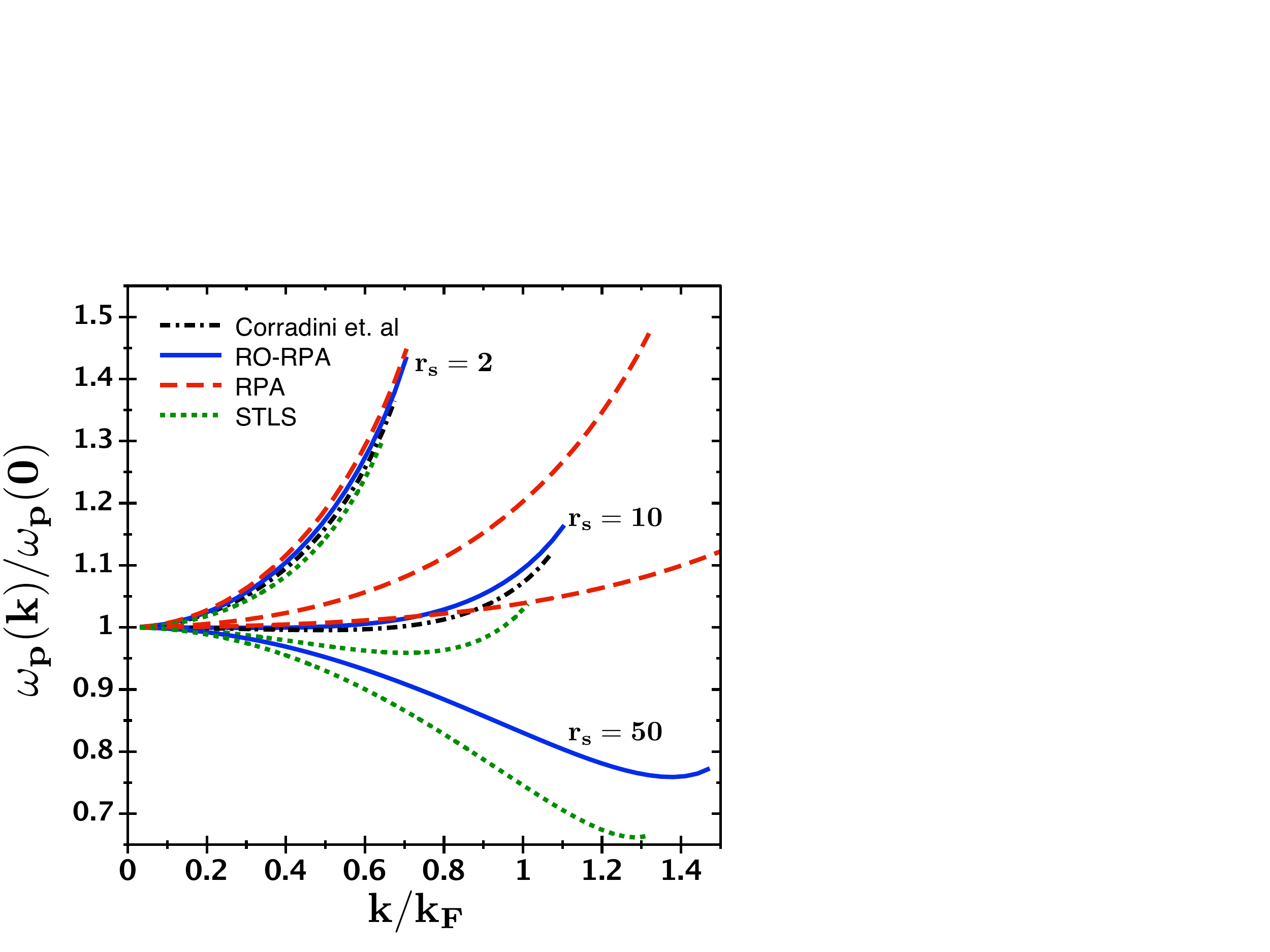}
\caption{\label{fig:plasmon}
Plasmon dispersion curves for various scaled average electron separations $r_s$.
The meanings of the curves are shown in the figure legend.
An $r_s$ label refers to the closest RO-RPA curve. 
For the other theories, the trend is for the slope of $\omega_p(k)$ at small $k$
to decrease as $r_s$ increases. The complete RPA curve for $r_s=50$ is not
shown, it terminating with value $\omega_p(k)/\omega_p(0) = 1.55$ at $k/k_F = 2.36$. The Corradini et al. curves were obtained using their fit\cite{corradini98}
of $G_+(k,0)$ to simulation data\cite{moroni95}.
}
\end{figure}
\begin{figure}
\includegraphics[scale=0.59]{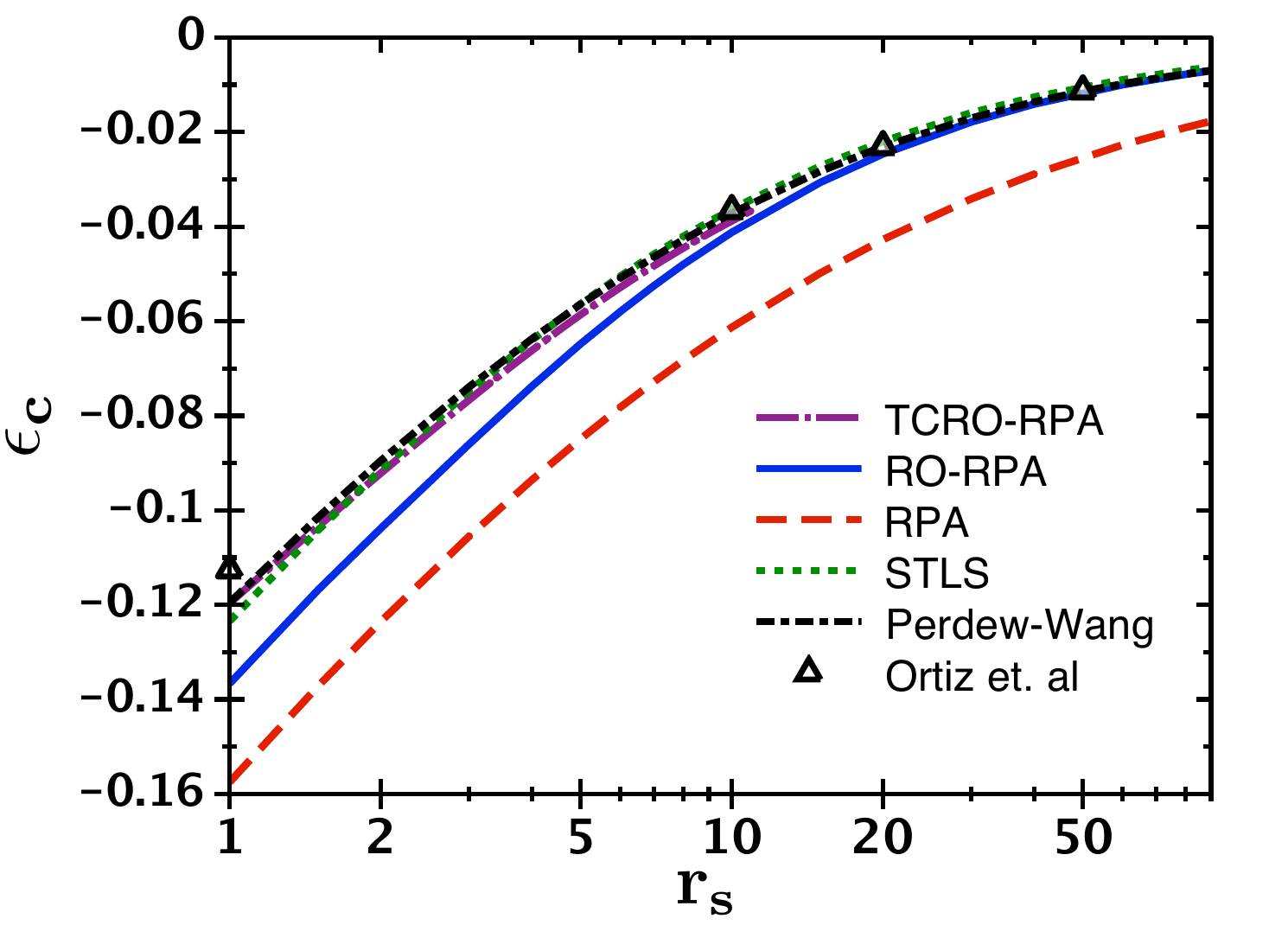}
\caption{\label{fig:ecorr}
Scaled correlation energy per particle $\epsilon_c$ (in Rydbergs) as a
function of the scaled average electron separation $r_s$.
The meanings of the curves and symbols are shown in the figure legend. The
Perdew-Wang curve is a fit \cite{perdew92} to simulation data \cite{ceperley80}.
}
\end{figure}

\begin{figure}
\includegraphics[scale=0.59]{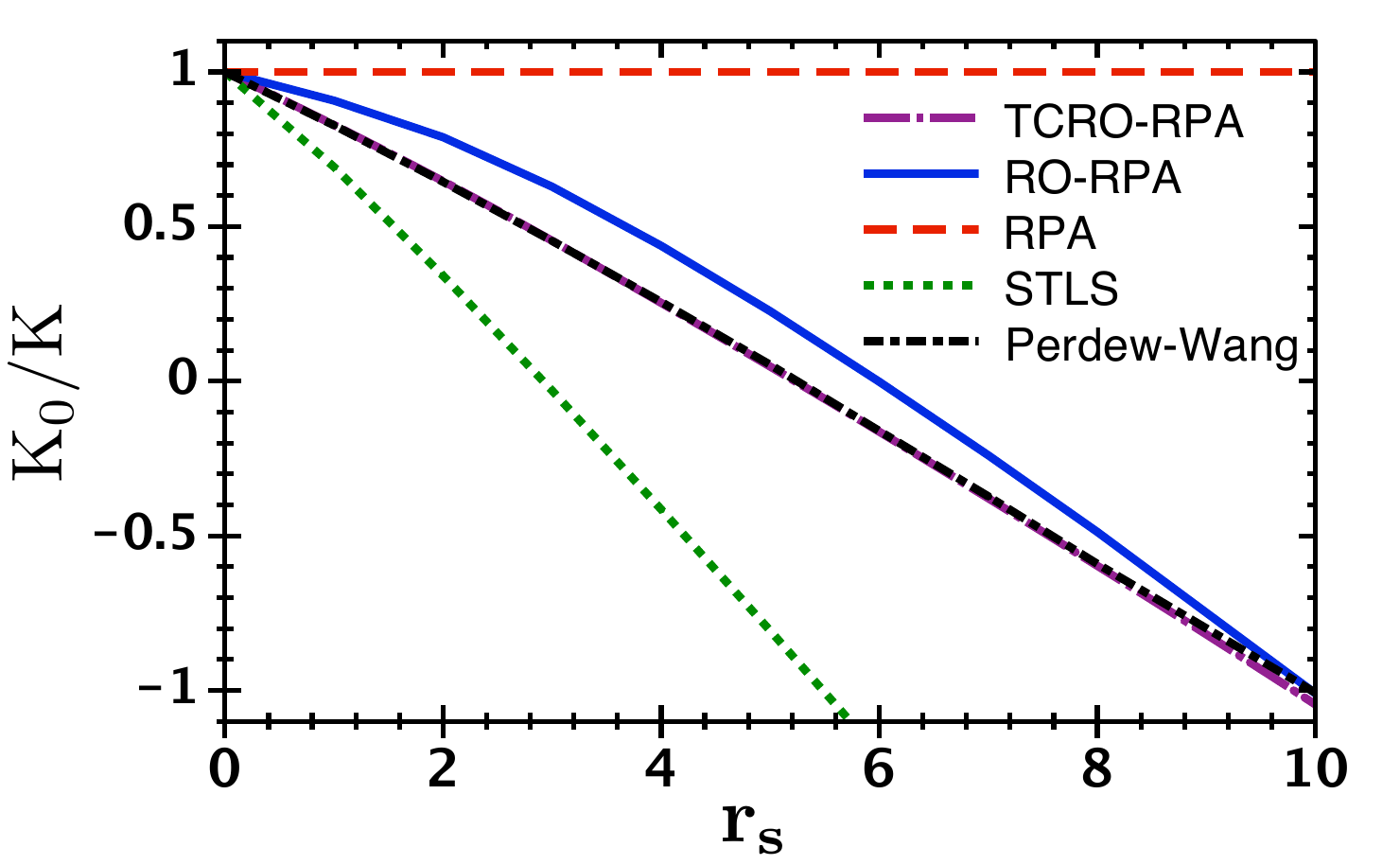}
\caption{\label{fig:kappainv}
Scaled inverse compressibility $K_0/K$ as a function of the scaled
average electron separation $r_s$, obtained using Eq.(\ref{compresseq}).
The meanings of the curves are shown in the figure legend. Note that the 
TCRO-RPA and Perdew-Wang curves overlap for almost all densities shown.
}
\end{figure}

\begin{figure}
\includegraphics[scale=0.59]{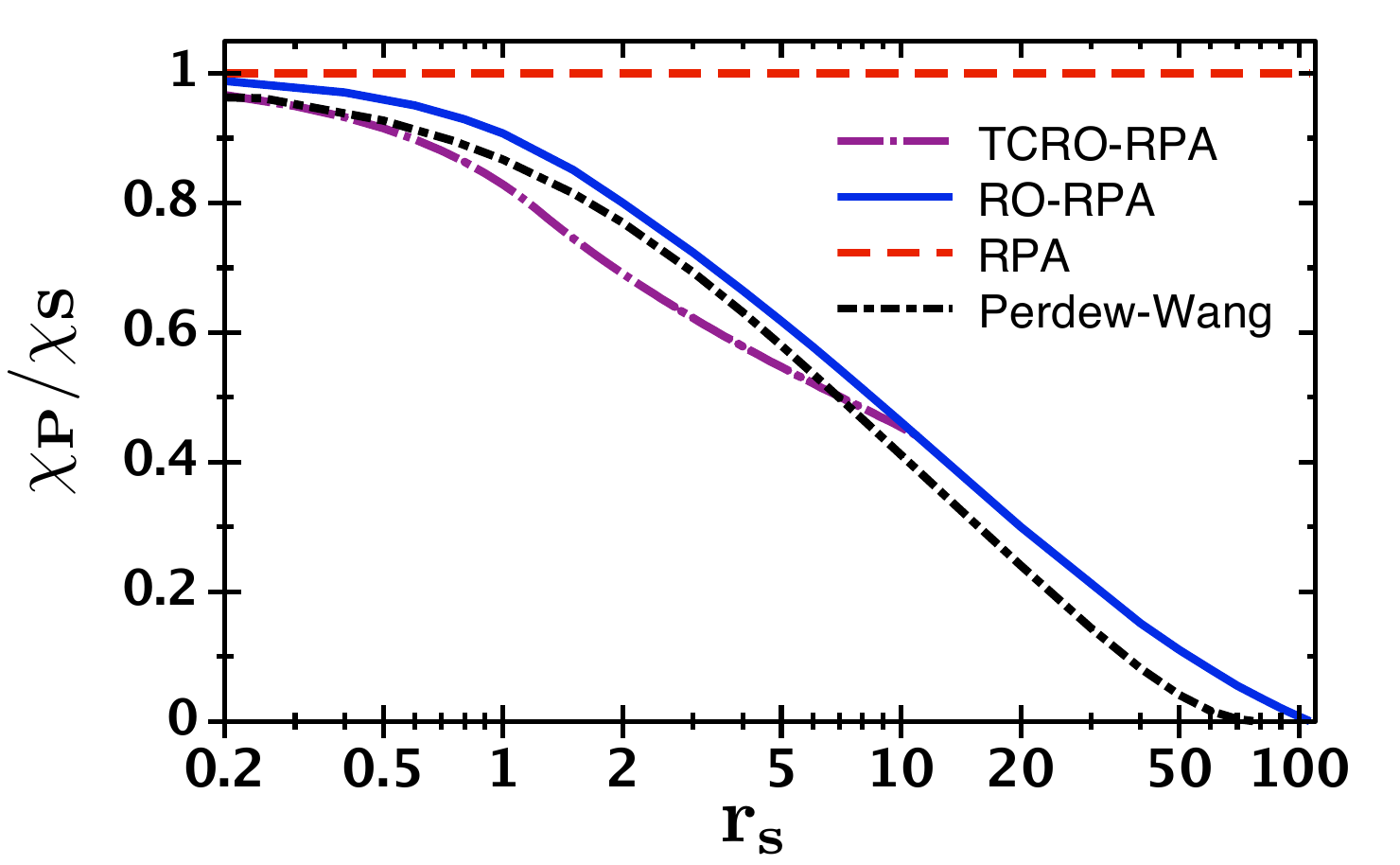}
\caption{\label{fig:chisinv}
Scaled inverse spin susceptibility $\chi_P/\chi_S$ as a function of the scaled 
average electron separation $r_s$, obtained using Eq.(\ref{suscepteq}).
The meanings of the curves are shown in the figure legend.
}
\end{figure}

As mentioned above, the focus on static properties partly justified ignoring the
frequency dependence of the optimized potentials, $u_{ij}$. However, the
structure of the multi-component theory is such that when mapped to the
one-component form for $\chi(k,\omega)$, Eq.(\ref{gfactoreq}), the local
field factor $G_+$ that arises is frequency dependent. It is
interesting then to examine theoretical predictions for a dynamic property of
the gas: its collective excitations, i.e., plasmons.\cite{fetter03,giuliani05} 

Figure \ref{fig:plasmon} shows the plasmon frequency $\omega_p$ as a function
of wavevector $k$ for various theories and simulation data for the same
densities as given in Figure \ref{fig:gr}. As shown, the plasmon frequency is
normalized by its value at zero wavenumber: $\omega_p(0) = \sqrt{4\pi ne^2/m}$.
The curves terminate at the beginning of the electron-hole continuum of the
non-interacting electron gas.\cite{giuliani05} The Corradini et al. curves
were obtained using their fit\cite{corradini98,giuliani05} of $G_+(k,0)$ to 
QMC simulation data of Moroni et al.\cite{moroni95} for $r_s=2$, 5 and 10.
While not shown in the figure, the predictions of the theory of
UI\cite{utsumi80} for $r_s=2$ are essentially the same as those shown for
Corradini et al.
As can be seen, the RO-RPA predicts larger plasmon frequencies than
the STLS theory for all densities, with the difference increasing
somewhat as $r_s$ increases. The RO-RPA predictions agree well with the results
using the Corradini et al. fit for both densities, $r_s=2$ and 10. 
It has been shown elsewhere that the fully self-consistent GWA gives poor predictions
for the spectral properties of jellium, including the plasmon modes.\cite{holm98}

Figure \ref{fig:ecorr} shows theoretical predictions for the scaled
correlation energy per particle, $\epsilon_c$, as a function of the
scaled average electron separation, $r_s$. Also shown are simulation data of
Ortiz et al.\cite{ortiz99}, and the Perdew-Wang fit\cite{perdew92} to
simulation data of Ceperley and Alder\cite{ceperley80}.
Vosko et al.\cite{vosko80} show results for $\epsilon_c$ for other
theories for the paramagnetic phase.
As can be seen, the predictions of the RO-RPA greatly improve upon those of the
RPA. As mentioned above, the density range of most metals is
$2\leq r_s\leq 6$.\cite{ashcroft76}
For this range, the RO-RPA values for $\epsilon_c$ are more negative than
those of simulation by 15\% on average, while the RPA values are more negative
by 46\% on average. The accuracy of the RO-RPA increases with
decreasing density such that, for example, at $r_s=40$, its value for
$\epsilon_c$ is within 3\% of simulation. The thermodynamically consistent
theory, TCRO-RPA, improves even more on the RPA: its predictions for
$\epsilon_c$ are within 4\% of simulation over the metallic density 
range. Interestingly, the predictions of the {\it one}-component RO-RPA theory,
while being inferior to the multi-component theory for almost every other
quantity, are slightly better than the multi-component for $\epsilon_c$ when
thermodynamic consistency is enforced. The one-component TCRO-RPA theory agrees with the Perdew-Wang fit to within 3\% for the metallic density range. This accuracy is the same or only slightly less than that of
the most accurate theories for $\epsilon_c$: STLS,\cite{stls68} and Vashishta
and Singwi,\cite{vashishta72} which agree with simulation within 2\% and 3\%,
respectively, over this density range (the UI theory agrees within 7\%\cite{utsumi80}). While it has been tested to date for only a few densities,
the fully self-consistent GWA gives very good agreement for
$\epsilon_c$, within 1\% for $r_s=2$ and 4.\cite{holm98} Given the mediocre
predictions of the theory for other properties, this good agreement is thought
to be to due to a large cancellation of effects.\cite{holm04}

Figure \ref{fig:kappainv} shows results for the scaled inverse of the
compressibility, $K_0/K$, as defined by the structure equation,
Eq.(\ref{compresseq}) above. Simulation values were obtained using
the Perdew-Wang fit\cite{perdew92} to data of Ceperley and
Alder\cite{ceperley80} for the correlation
energy, and an expression relating the compressibility to the total
energy\cite{giuliani05}. For jellium, simulation studies\cite{ceperley80}
show that the inverse compressibility goes
to zero at $r_s = 5.25$. The RO-RPA predicts a zero at $r_s = 6$,
which is $14\%$ higher. As a comparison, the STLS theory predicts a zero at
$r_s = 3$, which is 55\% less than simulation.
As is well known, the RPA gives the non-interacting value at all densities.
Interestingly, the TCRO-RPA predictions agree very well with the Perdew-Wang
fitted data, within 0.1\%, up to $r_s=8$.

Figure \ref{fig:chisinv} shows results for the scaled inverse of the
spin susceptibility, $\chi_P/\chi_S$, as defined by the structure
equation, Eq.(\ref{suscepteq}) above. As for $K$, simulation values
were obtained using the Perdew-Wang fit\cite{perdew92} for
the correlation energy, and an expression relating $\chi_S$
to the total energy\cite{giuliani05}. A divergence in $\chi_S$ is 
identified with a second-order transition from the paramagnetic state to
a polarized state. In Hartree-Fock theory, this polarized state was identified
with the fully polarized, i.e., ferromagnetic, state.\cite{fetter03}
Simulation work appeared to have reinforced this idea.\cite{ceperley80}
In the Hartree-Fock and RPA theories, the density of the divergence of 
$\chi_S$ is slightly lower (higher $r_s$) than the density for
paramagnetic-ferromagnetic phase coexistence.
In their fit then, Perdew and Wang reasoned that the true density of
divergence occurs at an $r_s$ slightly above that for paramagnetic-ferromagnetic coexistence, $r_s=73$, which was computed via simulation by Ceperley and
Alder.\cite{ceperley80}
This reasoning yielded a divergence of $\chi_S$ at $r_s=77.5$. This zero of $1/\chi_S$ can be seen in Figure \ref{fig:chisinv}.

Subsequent simulation work by Ortiz et al. found that partially polarized states
were energetically favorable at higher densities (lower $r_s$), and the
transition from the paramagnetic to a partially polarized state was
continuous.\cite{ortiz99} Later simulation work by Zong et al. underscored this
picture, though their estimate of the density of this transition was lower,
$r_s\approx 50$.\cite{zong02} In addition, Zong et al. estimated the transition
to the ferromagnetic state was at $r_s\approx 100$, which was also their
estimate for the transition to the Wigner solid state.\cite{zong02}

As can be seen in Figure \ref{fig:chisinv}, the
RO-RPA predicts a divergence at $r_s\approx 107$, in good agreement with the
estimate of Zong et al. for the paramagnetic-ferromagnetic transition.
So, either the RO-RPA overestimates the value of $r_s$ for the paramagnetic-partially polarized transition by a factor of two, or predicts that only the 
paramagnetic-ferromagnetic transition is second order. Given the accuracy
of the RO-RPA for the correlation energy at low density, the latter explanation
appears more plausible. However, the RO-RPA value for the density of divergence
was obtained from the structure route expression for the susceptibility, and
usually estimates using the energy route are more accurate.
Given the from of the RO-RPA multi-component theory, it is straightforward to
generalize it to compute $E_{corr}$ as a function of the fractional spin
polarization, $p \equiv \langle n_\uparrow(x)-n_\downarrow(x)\rangle/n$, and
thus determine coexistence and spinodal boundaries. The solution to this puzzle then is left to future research.

In this work, thermodynamic consistency was enforced on the compressibility,
which is a measure of the sensitivity of the electron density to changes in the pressure.
It is not expected then that this method would necessarily improve the
predictions of the theory for the spin
susceptibility, which is a measure of the sensitivity of a different quantity, the spin
polarization, to changes in the magnetic field.
Nonetheless, it is interesting to examine the TCRO-RPA predictions for $\chi_S$.
As can be seen in Figure \ref{fig:chisinv}, the TCRO-RPA values agree well
with the Perdew-Wang fit at very high density (small $r_s$), but drop below
the Perdew-Wang curve at $r_s\approx 0.5$. For example, at $r_s=2$, the
TCRO-RPA value is 10\% below the Perdew-Wang one. For $r_s>2$, the
TCRO-RPA curve asymptotically approaches the RO-RPA curve, terminating at
it at $r_s\approx 10.8$. At that point, the TCRO-RPA curve
is above the Perdew-Wang curve.
In other words, the agreement with the
fitted simulation data appears to be better for the RO-RPA than for
the TCRO-RPA. It has been remarked elsewhere that the Perdew-Wang fit is
not that accurate for polarized states.\cite{giuliani05} It seems unlikely
though that this inaccuracy would be that large near $p=0$. A clarifying task
then would be to compute
$\chi_S$ via the energy route. For this, $E_{corr}$ would be computed for
small $p$ and constant $r_s$, while enforcing thermodynamic
consistency on $K$ (or even $\chi_S$).

\section{\label{sec:summary}Summary and discussion}
In summary, the range optimization scheme was applied to the RPA theory
for jellium. It was shown that this RO-RPA theory gives greatly improved
predictions for the gas properties as shown by its
results for the pair correlation function $g(r)$, compressibility $K$ and spin
susceptibility $\chi_S$. For the correlation energy, $E_{corr}$, the theory is
most accurate at low densities, but it still gives predictions within 15\% of
simulation for the density range of most metals, $2\leq r_s\leq 6$. Enforcing
thermodynamic consistency on the compressibility improves the agreement with
simulation to within 4\% for this density range, while the one-component
version of the theory is slightly better at 3\%. This agreement is comparable
to the most accurate of the previous theories for
$E_{corr}$.\cite{stls68,vashishta72,holm98}
Also, the RO-RPA appears to outperform the STLS theory in
comparison with simulation data for the plasmon modes and compressibility,
and the pair correlation function at low density.

The thermodynamically consistent theory could be further improved by
conducting the range optimization to obey the cusp condition on
$g(r)$ near $r=0$.\cite{kimball73} Also, since range optimization can be applied
to any theory in which the positivity condition of $g(r)$ is violated,
improvements in the accuracy of the basic theory are possible.

One noteworthy result was that the RO-RPA theory predicts a divergence of
$\chi_S$ at a density that is lower than current estimates
for the liquid-solid transition.\cite{zong02,giuliani05} Thus, no
divergence of $\chi_S$ is predicted for the liquid phase. This result was
obtained using the structure route expression for
$\chi_S$, Eq.(\ref{suscepteq}).
Given the evidence from simulations for a transition to partially
polarized states,\cite{ortiz99,zong02} it would be interesting within
the RO-RPA to examine the phase behavior of jellium as a function of
polarization using the energy route.\cite{tanaka89} Further, since the temperature dependent density-density response function can be represented by an equation
almost identical to Eq.(\ref{chiinverse}),\cite{fetter03} the full temperature
dependent phase diagram can be obtained.

As stated above, one aim of this work is to apply range optimization to 
inhomogeneous electron liquids, to compute the band structure of semiconductors,
for example. It was shown in Sec.\ref{sec:numerical} how the new algorithm
to implement range optimization can be applied to inhomogeneous liquids, and
results for that will be presented in a future work.

\begin{acknowledgments}
I thank Craig Pryor for helpful conversations.
\end{acknowledgments}

\bibliography{paper}
\end{document}